\documentstyle[prd,aps]{revtex}

\newcommand{\bea}{\begin{eqnarray}}
\newcommand{\eea}{\end{eqnarray}}

\begin{document}

\draft
\twocolumn[\hsize\textwidth\columnwidth\hsize\csname
@twocolumnfalse\endcsname

\title{{\it COBE} constraints on inflation models with 
       a massive non-minimal scalar field}
\author{Jai-chan Hwang${}^{(a)}$ and Hyerim Noh${}^{(b)}$}
\address{${}^{(a)}$ Department of Astronomy and Atmospheric Sciences,
                    Kyungpook National University, Taegu, Korea \\
         ${}^{(b)}$ Korea Astronomy Observatory,
                    San 36-1, Whaam-dong, Yusung-gu, Daejon, Korea \\
        }
\date{\today}
\maketitle

\begin{abstract}
 
We derive power spectra of the scalar- and tensor-type structures
generated in an inflation model based on a massive non-minimally 
coupled scalar field with the strong coupling assumption.
We make analyses in both the original-frame and the conformally
transformed Einstein-frame.
We derive contributions of both structures to the anisotropy of the cosmic 
microwave background radiation, and compare the contributions with the 
four-year {\it COBE}-DMR data.
Previous study showed that sufficient amount of inflation requires
a small coupling parameter.
In such a case the spectra become near Zeldovich spectra, 
and the gravitational wave contribution becomes
negligible compared with the scalar-type contribution
which is testable in future CMBR experiments.

\end{abstract}

\noindent
\pacs{PACS number(s): 98.70.Vc, 98.80.Cq}

\vskip2pc]

{\it 1. Introduction:}
Inflation scenario, although still in hypothetical stage, 
now seems to have its firm presence in the theoretical
study of the early universe preceding the radiation dominated big bang era.
The main necessity of the inflation stems from the natural mechanism 
it provides concerning the origin of the large scale structures 
in the paradigm of spatially homogeneous and isotropic world model.
If one accepts the presence of inflation and its ability to 
generate seeds for the large scale structures, one can use the data of 
the observed structures to constrain the model parameters for the 
inflation models.
These days, the trouble is not the lack of plausible inflation
models but the presence of too many of them mostly based on toy models 
which are specially designed for the successful inflation and some based on 
hopeful theories for the physics in high-energy regime,
{\it i.e.,} the lack of standard model \cite{Turner}. 
However, often the observational constraints are strong enough to exclude 
models based on some promising high-energy physics; like the ones based on 
the grand unification \cite{GUT}, the supersymmetry \cite{SUSY},
the low-energy effective action of superstring theory \cite{PBB}, etc.
Although there are attempts to constrain the model parameters of
simple models based on a single field potential in Einstein gravity
\cite{design-inflation},
in the present situation without agreeable theory of the high-energy physics, 
the other common trend is to investigate the proposed inflation models
case by case, to constrain or to exclude them using the structure formation
processes and the observational data.

Recently, we notice growing interests on the roles of generalized versions
of gravity theories in reconstructing the early universe scenario.
These other-than Einstein gravity theories naturally arise either from 
attempts to quantize gravity \cite{Quantum-correction} or as the low-energy 
limits of the unified theories including gravity \cite{Unification}.
Naturally, assuming that the early universe was dominated by some of these
gravity theories, many inflation models have been proposed based on 
the generalized gravity theories \cite{Starobinsky,GGT-inflation}; 
in fact, the first physically motivated 
inflation model is the one by Starobinsky \cite{Starobinsky} based on 
the quantum corrected $R^2$ gravity which precedes the well credited 
inflation proposal by Guth \cite{Guth}.
There also have been many studies of the structure formation aspects
of such inflation models \cite{structure-GGT}.
We have been investigating the structure formation
processes in generalized gravity theories, and presented a unified way of
analyzing quantum generation and classical evolution processes of the 
scalar- and tensor-type structures applicable in a wide class of
generalized gravity theories \cite{QG-CE,GGT-scalar,GGT-GW}, 
and made applications to string gravity \cite{PBB} and 
nonminimally coupled scalar field with a self coupling \cite{NM-1}.
In this paper we will make another application to a chaotic type
inflation model based on a nonminimally coupled massive scalar field
proposed by Futamase and Maeda \cite{FM} and will derive constraints 
on the model using the recent {\it COBE}-DMR data.
The main results are in Eqs. (\ref{a_l},\ref{a_l-CT}) and
(\ref{limit},\ref{limit-CT}), and discussions below Eq. (\ref{limit-CT}).

\vskip .5cm
{\it 2. Gravity theories:}
We consider the following form of generalized gravity theory
\bea
   {\cal L} = \sqrt{-g}
       \left[ {1 \over 2} f (\phi, R)
       - {1 \over 2} \omega (\phi) \phi^{;c} \phi_{,c}
       - V (\phi) \right]. 
   \label{Lagrangian}
\eea
Unified analyses of the quantum generation and the classical evolution 
processes of the scalar- and tensor-type structures based on this 
gravity theory were presented in \cite{GGT-scalar,GGT-GW}. 
A non-minimally coupled massive scalar field is a case with 
$f = (\kappa^{-2} - \xi \phi^2)R$, $\omega = 1$, and 
$V = {1 \over 2} m^2 \phi^2$ where $\kappa^2 \equiv 8 \pi m_{pl}^{-2}$;
we call it the {\it original-frame}.
Through a conformal transformation \cite{GGT-CT-orig,GGT-CT} 
this gravity theory can be transformed to Einstein gravity 
($f = \kappa^{-2} R$, $\omega = 1$) with a special potential, 
which is also a special case of Eq. (\ref{Lagrangian});
we call it the {\it Einstein-frame}.
Since the general results concerning cosmic structures based on
Eq. (\ref{Lagrangian}) were presented in \cite{GGT-scalar,GGT-GW}, 
in the following we will use those results freely.

\vskip .5cm
{\it 3. The original-frame:}
Equations for the background are in Eq. (5) of \cite{GGT-scalar}.
{\it Assuming} the strong coupling condition $\kappa^2 |\xi| \phi^2 \gg 1$ 
we have the following solutions for the background 
(we consider $\xi < 0$ case) \cite{FM,Sakai}:
\bea
   & & a \propto e^{H_i t}, \quad
       H_i \equiv { (1 - 4 \xi ) m \over \sqrt{ -2 \xi ( 1 - 6 \xi )
       ( 3 - 16 \xi ) } }, 
   \nonumber \\
   & & 
       \phi \propto e^{\alpha m t}, \quad
       \alpha \equiv \sqrt{ - 2 \xi \over ( 1 - 6 \xi ) ( 3 - 16 \xi ) }.
   \label{BG}
\eea
Thus, we have a near exponentially expanding period which can provide 
a plausible inflationary era.

The on-shell Lagrangians for the scalar- and tensor-type structures
are presented in Eq. (7) of \cite{GGT-scalar}
and Eq. (3,24) of \cite{GGT-GW}, respectively.
The background evolution in Eq. (\ref{BG}) leads to 
\bea
   \nu = \nu_g = {3 - 16 \xi \over 2 ( 1 - 4 \xi )},
   \label{nu}
\eea
which are defined in Eqs. (14,21) of \cite{GGT-scalar} 
and Eqs. (8,11) of \cite{GGT-GW}.
Thus, the {\it general} mode function solutions in Eq. (14) of \cite{GGT-scalar}
and Eq. (29) of \cite{GGT-GW} include our case as a special subset.
In the large-scale limit, the general power spectra
based on vacuum expectation values in Eq. (16) of \cite{GGT-scalar} 
and Eq. (32) of \cite{GGT-GW} lead to the following:
\bea
   {\cal P}_{\hat \varphi_{\delta \phi}}^{1/2} 
   &=& {H \over |\dot \phi|} {\cal P}_{\delta \hat \phi_\varphi}^{1/2} 
       = {H^2 \over 2 \pi |\dot \phi|} {\Gamma(\nu) \over \Gamma(3/2)}
       {\sqrt{1 - 6 \xi} \over 1 - 4 \xi}
       \left( { k |\eta| \over 2 } \right)^{3/2-\nu}
   \nonumber \\
   & & \times \big| c_2 (k) - c_1 (k) \big|,
   \label{P-s-NM-q} \\
   {\cal P}_{\hat C_{\alpha\beta}}^{1/2} 
   &=& {\kappa H \over \sqrt{2} \pi} {1 \over \sqrt{1 - \kappa^2 \xi \phi^2} }
       {\Gamma(\nu_g) \over \Gamma(3/2)}
       \left( { k |\eta| \over 2 } \right)^{3/2-\nu_g}
   \nonumber \\
   & & \times
       \sqrt{ {1 \over 2} \sum_\ell 
       \big| c_{\ell 2} (k) - c_{\ell 1} (k) \big|^2 },
   \label{P-GW-NM-q}
\eea
where $\ell$ indicates the two polarization states of the gravitational wave.
In these forms, $\xi = 0$ reproduces correctly the minimally coupled limit;
see Eqs. (56,57,6) of \cite{GGT-scalar} and Eqs. (34,37) of \cite{GGT-GW}. 
$c_i(k)$ and $c_{\ell i}(k)$ are constrained by the quantization 
conditions $|c_2|^2 - |c_1|^2 = 1$, and $|c_{\ell 2}|^2 - |c_{\ell 1}|^2 = 1$.
Notice the general dependences of the power spectra on the scale $k$
through the vacuum choices which fix $c_i$ and $c_{\ell i}$.
Only if we choose the simplest vacuum states $c_2 = 1$ and $c_{\ell 2} = 1$
the power spectra show power-law dependence on $k$.

As an {\it ansatz} we {\it identify} the power spectra based on the vacuum 
expectation value during the inflation era 
[${\cal P}_{\hat \varphi_{\delta \phi}}$ and 
${\cal P}_{\hat C_{\alpha\beta}}$]
with the classical power spectra based on the spatial average
[${\cal P}_{\varphi_{\delta \phi}}$ and ${\cal P}_{C_{\alpha\beta}}$];
by this identification we have ignored the possible contributions from 
classicalization processes.
Then, we have the same results in 
Eqs. (\ref{P-s-NM-q},\ref{P-GW-NM-q}) now valid for 
${\cal P}_{\varphi_{\delta \phi}}$ and ${\cal P}_{C_{\alpha\beta}}$.
In \cite{GGT-scalar,GGT-GW} we have shown that, ignoring the transient 
solutions, $\varphi_{\delta \phi}$ and $C_{\alpha\beta}$ are generally
{\it conserved} independently of changing gravity [from non-minimally coupled 
to minimally coupled], changing potential, and changing equation of state, 
as long as the scale remains in the large-scale; this is the case for the
observationally relevant scales before the second (inward) horizon crossing
in the matter dominated era.
Consequently, even in the matter dominated era 
Eqs. (\ref{P-s-NM-q},\ref{P-GW-NM-q})
remain valid as the power spectra in the large-scale limits 
[the super-sound-horizon for the scalar-type structure 
and the super-horizon for the gravitational wave].
The spectral indices of the scalar- and tensor-type structures are
given as $n -1 = 3 - 2 \nu$ and $n_g = 3 - 2 \nu_g$; see below
Eq. (60) of \cite{GGT-scalar} and below Eq. (52) of \cite{GGT-GW}.
Thus, {\it ignoring} the vacuum dependences in the inflation era
(i.e., taking the simplest vacuum state), we have
\bea
   n - 1 = n_g = {4 \xi \over 1 - 4 \xi},
   \label{indices}
\eea
which is less than zero, thus showing redder spectra compared with the
scale independent ones ($n-1 = 0 = n_g$); 
results up to this point are valid for general negative value of $\xi$.
In order to be consistent with the {\it COBE} data with 
$n-1 \simeq 0 \simeq n_g$ we need $|\xi| \ll 1$.
In this limit we have:
\bea
   & & {\cal P}_{\varphi_{\delta \phi}}^{1/2}
       = \sqrt{ 3 \over - 2 \xi} {H_i^2 \over 2 \pi m \phi}, \quad
       {\cal P}_{C_{\alpha\beta}}^{1/2}
       = \sqrt{ 2 \over - \xi} {H_i \over 2 \pi \phi}. 
   \label{P}
\eea

In order to confront the theory with the observational data we derive the 
contribution to angular power-spectrum of the cosmic microwave
background radiation.
The multipole $\langle a_l^2 \rangle \equiv \langle |a_{lm}|^2 \rangle$ 
is related to ${\cal P}_{\varphi_{\delta \phi}}$
and ${\cal P}_{C_{\alpha\beta}}$ through formulae
in Eq. (61) of \cite{GGT-scalar} and Eq. (56) of \cite{GGT-GW}.
Thus, the observed values (or limits) of $\langle a_l^2 \rangle$ constrain 
directly ${\cal P}_{\varphi_{\delta \phi}}$ and ${\cal P}_{C_{\alpha\beta}}$,
thus constrain free parameters in the model.

{}For the scale independent spectra in Eq. (\ref{P}) $\langle a_l^2 \rangle$ 
can be integrated \cite{NM-1}.
The quadrupole anisotropy is the following
\bea
   \langle a_2^2 \rangle 
   &=& \langle a_2^2 \rangle_S + \langle a_2^2 \rangle_T
   \nonumber \\
   &=& {\pi \over 75} {\cal P}_{\varphi_{\delta \phi}}
       + 7.74 {1 \over 5} {3 \over 32} {\cal P}_{C_{\alpha\beta}}
   \nonumber \\
   &=& {1 \over 480 \pi \xi^2} \left( {m \over \phi} \right)^2
       \left( {1 \over - 15 \xi} 
       + {3 \times 7.74 \over 4 \pi} \right).
   \label{a_l}
\eea
Thus, if $|\xi| < 1/27.7$ the gravitational wave contribution
is smaller compared with the scalar-type one.
The four-year {\it COBE}-DMR data shows \cite{COBE}:
\bea
   & & Q_{\rm rms-PS} = 18 \pm 1.6 \mu K, \quad
       T_0 = 2.725 \pm 0.020 K,
   \nonumber \\
   & & \langle a_2^2 \rangle 
       = {4 \pi \over 5} \left( {Q_{\rm rms-PS} \over T_0} \right)^2 
       \simeq 1.1 \times 10^{-10}.
   \label{a_l-value}
\eea
This leads to a constraint
\bea
   {m \over \phi} \simeq 1.6 \times 10^{-3} |\xi|
       \sqrt{ |\xi| \over 1 + 27.7 |\xi| }.
   \label{limit}
\eea

\vskip .5cm
{\it 4. The Einstein-frame:}
Exactly similar analyses can be made in this gravity theory.
{\it Assuming} the strong coupling limit ($\kappa^2 |\xi| \phi^2 \gg 1$)
in the original-frame, the potential in the Einstein-frame shows an asymptotic 
form $V = {m^2 \over 2 \xi^2 \kappa^2} e^{-\sqrt{- 4 \xi \over 1 - 6 \xi}
\kappa \phi}$.
Due to the exponential potential, we have a power-law type accelerated
expansion \cite{power-law}:
\bea
   & & a \propto t^p, \quad
       p = 3 - {1 \over 2 \xi}, 
   \nonumber \\
   & & \kappa \phi = \sqrt{2 p} \ln{ \left[
       \sqrt{2 \over (1 - 6 \xi) ( 3 - 16 \xi)} m t \right] }.
   \label{BG-CT}
\eea
This leads to the same result in Eq. (\ref{nu}) for $\nu$ and $\nu_g$.
The large-scale power spectra become:
\bea
   {\cal P}_{\hat \varphi_{\delta \phi}}^{1/2} 
   &=& {H^2 \over 2 \pi |\dot \phi|} {\Gamma(\nu) \over \Gamma(3/2)}
       {1 - 4 \xi \over 1 - 6 \xi}
       \left( { k |\eta| \over 2 } \right)^{3/2-\nu}
   \nonumber \\
   & & \times \big| c_2 (k) - c_1 (k) \big|,
   \label{P-s-NM-q-CT} \\
   {\cal P}_{\hat C_{\alpha\beta}}^{1/2} 
   &=& {\kappa H \over \sqrt{2} \pi}
       {\Gamma(\nu_g) \over \Gamma(3/2)}
       {1 - 4 \xi \over 1 - 6 \xi}
       \left( { k |\eta| \over 2 } \right)^{3/2-\nu_g}
   \nonumber \\
   & & \times
       \sqrt{ {1 \over 2} \sum_\ell 
       \big| c_{\ell 2} (k) - c_{\ell 1} (k) \big|^2 },
   \label{P-GW-NM-q-CT}
\eea
with similar constraints on $c_i(k)$ and $c_{\ell i}(k)$ as below
Eq. (\ref{P-GW-NM-q}).
These results, of course, are consistent with Eq. (57) in \cite{GGT-scalar}
and Eq. (37) in \cite{GGT-GW}.

After {\it identifying} the power spectra based on the vacuum expectation 
values with the classical power spectra based on the spatial average,
and using the {\it conservation properties} in the large-scale limit,
we have the same results in 
Eqs. (\ref{P-s-NM-q-CT},\ref{P-GW-NM-q-CT}) now valid for 
${\cal P}_{\varphi_{\delta \phi}}$ and ${\cal P}_{C_{\alpha\beta}}$.
{\it Ignoring} the vacuum dependences in the inflation era we have the same 
result in Eq. (\ref{indices}) now valid for the spectral indices $n$ and $n_g$.
Thus, to be consistent with the {\it COBE} data, again we need $|\xi| \ll 1$.
In this limit we have:
\bea
   & & {\cal P}_{\varphi_{\delta \phi}}^{1/2}
       = {1 \over \sqrt{- 2 \pi \xi}} {H \over m_{pl}}, \quad
       {\cal P}_{C_{\alpha\beta}}^{1/2}
       = { 2 \over \sqrt{\pi}} {H \over m_{pl}}. 
   \label{P-CT}
\eea
{}For the scale independent spectra in Eq. (\ref{P-CT}) the quadrupole 
anisotropy becomes
\bea
   \langle a_2^2 \rangle 
       = {1 \over 10} \left( {H \over m_{pl}} \right)^2
       \left( {1 \over - 15 \xi} 
       + {3 \times 7.74 \over 4 \pi} \right).
   \label{a_l-CT}
\eea
Thus, for $|\xi| < 1/27.7$ the gravitational wave contribution
is smaller compared with the scalar-type one.
Using the four-year {\it COBE}-DMR data in Eq. (\ref{a_l-value}) we have 
a constraint
\bea
   {H \over m_{pl}} \simeq 1.3 \times 10^{-4} \sqrt{ |\xi| \over
        1 + 27.7 |\xi| }.
   \label{limit-CT}
\eea

Based on analyses in the Einstein frame, in order to have successful 
amount of inflation, authors of \cite{FM} showed $|\xi| < 10^{-3}$.
In such a case the spectra of both structures in Eq. (\ref{indices})
show near Zeldovich spectra which is consistent with the {\it COBE}-DMR data.
According to our above analyses
such a small coupling parameter leads to two important consequences.
Firstly, based on Eq. (\ref{a_l-CT}) it predicts that the contribution of 
the gravitational wave should be negligible compared with 
the scalar-type perturbation.
This is a testable result in future CMBR experiments like MAP
and Planck Surveyor missions with high accuracy
temperature and polarization anisotropies.
Secondly, based on Eq. (\ref{limit-CT}) we have a limit on the expansion
rate during inflation like $H < 4 \times 10^{-6} m_{pl}$.

\vskip .5cm
{\it 5. Discussions:}
In Sec {\it 2.} we mentioned about the relation between the two frames
through the conformal transformation \cite{CT-history}.
Using the conformal transformation properties in Eqs. (11-14) of \cite{GGT-CT} 
we can show that through the conformal transformation the original-frame 
results in Eqs. (\ref{P-s-NM-q},\ref{P-GW-NM-q},\ref{P},\ref{a_l})
produce correctly the Einstein-frame results in Eqs. 
(\ref{P-s-NM-q-CT},\ref{P-GW-NM-q-CT},\ref{P-CT},\ref{a_l-CT}).
In Eq. (25) of \cite{GGT-CT} we showed that $\varphi_{\delta \phi}$
and $C_{\alpha\beta}$ are invariant under the conformal transformation.
As far as we can tell, however, the two theories related by the 
conformal transformation are not necessarily equivalent physically; 
for interesting discussions, see \cite{CT}.
Still, as long as the structure generation and evolution processes based on
the linear theory are concerned, as noted above, the conformal transformation
provides a simple and useful {\it mathematical trick} 
for practical calculations.

In this paper we have derived the the scalar- and tensor-type 
structures generated from quantum fluctuations in a chaotic inflation model 
based on a massive non-minimally coupled scalar field.
We derived contributions of both structures to the anisotropy of the cosmic 
microwave background radiation and compared them with the four-year 
{\it COBE}-DMR data on quadrupole anisotropy.
The power spectra in 
Eqs. (\ref{P-s-NM-q},\ref{P-GW-NM-q},\ref{P-s-NM-q-CT},\ref{P-GW-NM-q-CT}) 
and their corresponding classical ones after classicalization and
subsequent conserved evolution, and the spectral indices in Eq. (\ref{indices}),
are valid for general negative value of $\xi$.
In order to investigate the constraints from the {\it COBE} data in such 
a general $\xi$ situation we can numerically integrate the general multipole 
formulae in Eq. (61) of \cite{GGT-scalar} and Eq. (56) of \cite{GGT-GW}.
{}For a small $|\xi|$ the spectra are nearly scale-independent and 
the results in the original frame and the Einstein frame are
presented in Eqs. (\ref{limit},\ref{limit-CT}), respectively.
Also, for very small $|\xi|$ the gravitational wave contribution becomes
negligible compared with the scalar-type contribution.
In fact, for a successful amount of inflation in the Einstein frame
the authors of \cite{FM} showed that $|\xi| < 10^{-3}$, and in such 
a case the gravitational wave should give negligible contribution
to the CMBR temperature anisotropy
which is a testable result in future CMBR experiments.
That is, an excessive amount of gravitational wave signature detected 
in the future CMBR experiments can rule out the inflation scenario 
proposed by Futamase and Maeda in \cite{FM}.
A way of detecting gravitational wave signature using a particular
combination of polarization parameters was discussed in \cite{CMBR-GW}.
Expected accuracy of temperature and polarization anisotropies 
in future CMBR experiments using MAP and Planck satellite missions 
are investigated in \cite{CMBR-forecasts}.

Recently, the Sakai and Yokoyama \cite{Sakai} proposed a scenario
based on a topological inflation \cite{topological-inflation}
realized in the domain wall formed at the maximum of the effective potential 
in the Einstein frame.
Analyzing the scalar and tensor-type contributions to CMBR in this
scenario would be an interesting subject which will be addressed
in a future occasion.

\vskip .5cm
JH wishes to thank Dr. Nobuyuki Sakai for drawing our attention to the
subject during a cosmology conference in the University of Cape Town
and for useful discussions.
We would like to thank the referee for informing us an important point
we missed in the original manuscript.


\end{document}